# Terahertz phonon engineering with van der Waals heterostructures

Yoseob Yoon[1,2,3,*], Zheyu Lu[2,4], Can Uzundal[2,5], Ruishi Qi[1,2], Wenyu Zhao[1], Sudi Chen[1,6], Qixin Feng[1,2], Woochang Kim[1,2], Mit H. Naik[1,2], Kenji Watanabe[7], Takashi Taniguchi[8], Steven G. Louie[1,2], Michael F. Crommie[1,2,6], and Feng Wang[1,2,6,*]

[1]Department of Physics, University of California, Berkeley, CA, USA

[2]Materials Sciences Division, Lawrence Berkeley National Laboratory, Berkeley, CA, USA

[3]Department of Mechanical and Industrial Engineering, Northeastern University, Boston, MA, USA

[4]Graduate Group in Applied Science and Technology, University of California, Berkeley, CA, USA

[5]Department of Chemistry, University of California, Berkeley, CA, USA

[6]Kavli Energy NanoScience Institute, Berkeley, CA, USA

[7]Research Center for Electronic and Optical Materials, National Institute for Materials Science, Tsukuba, Japan

[8]Research Center for Materials Nanoarchitectonics, National Institute for Materials Science, Tsukuba, Japan

[*]Correspondence to: y.yoon@northeastern.edu; fengwang76@berkeley.edu

# Abstract

Phononic engineering at gigahertz (GHz) frequencies form the foundation of microwave acoustic filters[1], acousto-optic modulators[2], and quantum transducers[3,4]. Terahertz (THz) phononic engineering could lead to acoustic filters and modulators at higher bandwidth and speed, as well as quantum circuits operating at higher temperatures. Despite its potential, methods for engineering THz phonons have been limited due to the challenges of achieving the required material control at sub-nanometer precision and efficient phonon coupling at THz frequencies. Here, we demonstrate efficient generation, detection, and manipulation of THz phonons through precise integration of atomically thin layers in van der Waals heterostructures. We employ few-layer graphene (FLG) as an ultrabroadband phonon transducer, converting femtosecond near-infrared pulses to acoustic phonon pulses with spectral content up to 3 THz. A monolayer $WSe_2$ is used as a sensor, where high-fidelity readout is enabled by the exciton-phonon coupling and strong light-matter interactions. Combining these capabilities in a single heterostructure and detecting responses to incident mechanical waves, we perform THz phononic spectroscopy. Using this platform, we demonstrate high-Q THz phononic cavities and show that a monolayer $WSe_2$ embedded in hexagonal boron nitride (hBN) can efficiently block the transmission of THz phonons. By comparing our measurements to a nanomechanical model, we obtain the force constants at the heterointerfaces. Our results could enable THz phononic metamaterials for ultrabroadband acoustic filters and modulators, and open novel routes for thermal engineering.

# Main

Engineering and control of acoustic phonons at GHz frequencies have enabled microwave acoustic filters and the transduction of qubits for quantum information processing. THz phononic engineering could lead to acoustic technology at higher speed and larger bandwidth, as well as single-phonon quantum states at higher temperatures. In addition, the thermal transport of nonmetallic solids is mostly governed by acoustic phonons at THz frequencies at room temperature (300 K ~ 26 meV ~ 6 THz). THz phononic engineering may therefore enable rational design of structured materials with optimal thermal properties.

Despite its potential, phonon engineering at THz frequencies has been little explored so far due to experimental challenges in the coherent generation, detection, and control of THz acoustic phonons. At THz frequencies, the phonon wavelengths are around 3 nm for a sound speed of 3 km/s. Consequently, coherent THz phonon generation and manipulation require highly precise material control at sub-nanometer or atomic precision. In addition, broadband detection of THz phonons requires not only ultrafast temporal responses but also ultrahigh sensitivity to vibrations in nanometer-thick materials.

To generate coherent acoustic phonons, piezoelectric materials or thin metals are commonly used as a transducer that converts electrical or optical signals to mechanical signals. Conventional piezoelectric transducers can generate coherent phonons at kHz to GHz frequencies, while thin metals excited by ultrafast optical pulses can generate coherent phonons up to 1 THz (refs. [5–7]). Multiple-quantum-well structures based on GaAs/AlAs or GaN/InGaN have enabled the generation of acoustic phonons up to 1.4 THz (refs. [8–13]), but it is achieved only at narrowband frequencies.

Van der Waals heterostructures provide a powerful material platform for THz phononic engineering. Many different van der Waals layers can be stacked into heterostructures with unit-cell thickness control and atomically smooth interfaces, which is ideal for coherent phonon generation and manipulation at THz frequencies. In addition, van der Waals materials such as graphene and transition metal dichalcogenides (TMDs) show femtosecond carrier dynamics and/or remarkably strong light-matter interactions, enabling the ultrafast transduction and highly sensitive detection of coherent phonons[14–19] and the fabrication of sub-THz phononic cavities[20].

Here, we demonstrate efficient broadband THz phonon generation up to 3 THz with FLG, sensitive THz phonon detection with monolayer WSe$_2$, and flexible THz manipulation with suitably designed van der Waals heterostructure stacks. By detecting responses to incident mechanical waves, we perform THz phononic spectroscopy, similar to conventional optical spectroscopy which detects responses to incident electromagnetic waves. We show that a single layer of WSe$_2$ embedded in hBN can efficiently block the transmission of THz phonons. We also quantitatively determine the acoustic phonon propagation speed in hBN and the force constants at the graphene-hBN and WSe$_2$-hBN heterointerfaces.

**Generation and detection of THz phonons**

We use van der Waals heterostructures composed of FLG, hBN, and WSe$_2$ for coherent generation, detection, and manipulation of THz acoustic phonons. Figure 1a illustrates a representative configuration of the heterostructure. We chose FLG for THz phonon transduction because its ultrafast carrier dynamics and light atomic mass facilitate broadband and high-frequency phonon generation. The FLG thickness $d_{\text{FLG}} = N_g d_g$ limits the characteristic acoustic phonon wavelength ($\lambda \sim 2d_{\text{FLG}}$) that can be launched by the transducer, where $N_g$ is the number of graphene layers and $d_g = 0.335$ nm is the graphene thickness. The central frequency ($f_c$) and bandwidth ($\Delta f$) of the generated phonons can be estimated[18,20] as $f_c \sim \Delta f \sim v_s/\lambda \sim v_s/2d_{\text{FLG}}$, where $v_s$ is the out-of-plane sound speed. Graphite has a large value of $v_s = 4.0$ km/s due to the light atomic mass of carbon, which enables the generation of THz phonons over 3 THz in the bilayer limit ($N_g = 2$). In contrast, the acoustic phonons generated at the bilayer TMD limit is less than 1 THz, limited by their heavy masses[18]. Graphene's ultrafast hot-carrier relaxation dynamics[21] and extreme thermal nonlinearity[22] also help the high-frequency acoustic phonon generation.

Experimentally, we use ultrafast pump-probe spectroscopy to generate and detect THz phonons. We tuned the pump photon energy to 1.409 eV so that the pump laser creates hot carriers in FLG but not in WSe$_2$ and hBN. The ultrafast thermal expansion and recoil of FLG (Extended Data Fig. 1) launch an acoustic phonon pulse (B$_{1g}$ mode) that propagates across hBN. Note that the full width at half maximum (FWHM) of pump and probe beams (10-15 μm and 1.6 μm, respectively) are much larger than THz phonon wavelengths, so the in-plane phonon modes are neither generated nor detected.

For broadband and sensitive detection of the coherent THz phonons, we choose monolayer WSe$_2$ as a sensing layer due to its strong light-matter interactions[23]. When a coherent phonon pulse reaches the WSe$_2$ layer, the WSe$_2$-hBN distance oscillates due to the nanomechanical motion. The expansion and compression of layers modulate the effective dielectric constant $\epsilon_{\text{eff}}$ around the WSe$_2$ layer, resulting in a shift of the WSe$_2$ exciton energy[24–26]. A 1 pm change in hBN-WSe$_2$ distance, for the equilibrium interlayer distance of around 400 pm, can shift the exciton energy by ∼50 μeV (ref. [26]). This effective coupling between acoustic phonons and excitons allows us to detect phonon oscillations by monitoring the exciton energy using resonant optical probe pulses. Even with a monolayer sensor, which is required for THz bandwidth, a high-fidelity readout is possible with the excitonic effect in WSe$_2$. Using this differential spectroscopic readout of the WSe$_2$ exciton peak, we are able to distinguish the phonon pulses at the WSe$_2$ layer from other signals from the FLG reflectivity change or air-gap modulation; it is difficult to distinguish them with other techniques.

Figure 1b shows coherent phonon-induced exciton oscillations in a WSe$_2$/hBN/3LG heterostructure (Device 1). The pump-induced transient reflectance change, defined as

$$\Delta R/R \equiv \frac{R_{\text{pump on}} - R_{\text{pump off}}}{R_{\text{pump off}}} \quad (1)$$

is plotted as a function of the pump-probe time delay ($\Delta t$) and probe energy, where $R_{\text{pump on}}$ and $R_{\text{pump off}}$ are probe reflectance spectra with and without the pump beam, respectively. From the absolute value and line shape of $\Delta R/R$ spectrum at each $\Delta t$, the pump-induced exciton energy shift can be estimated, similar to the transfer-matrix calculation performed in ref. [27]. Typical values of phonon-induced exciton shifts are up to 300 μeV (Extended Data Fig. 2) at pump fluence of 250 μJ/cm$^2$, which corresponds to the hBN layer displacement (phonon amplitude) of 6 pm. This linearly increases with the pump fluence and saturates at a higher pump fluence of ∼1 mJ/cm$^2$ (Extended Data Fig. 3). This imposes the limit for the THz phonon amplitude; a superlattice of FLG/hBN/FLG/hBN… would overcome this limit and produce higher phonon amplitudes at THz frequencies.

We used a Hann window with FWHM of 0.5 THz (−0.25 to 0.25 THz) to filter out the quasi-DC component and show the oscillating component in Fig. 1c. A line cut is obtained by calculating the difference between the high-energy (A) and low-energy (B) sides of the exciton peak, as shown in Fig. 1d. A Fourier transform of the time trace shows the spectral content of generated acoustic phonons (Fig. 1e), which spans up to 3 THz with a bandwidth (FWHM) of 1.2 THz. These are the highest values for coherently generated broadband acoustic phonons to our knowledge. Such high-frequency and broadband phonons are extremely useful for performing THz phononic spectroscopy, as we show below.

## High-quality THz phononic cavity

We first use THz phononic spectroscopy to quantitatively determine the acoustic phonon propagation speed $v_s$. Although $v_s$ can be estimated from phonon dispersion obtained by inelastic X-ray scattering[28], its energy resolution is limited for the accurate determination of $v_s$. Furthermore, the out-of-plane phonon mode of hBN is silent in Raman measurements[29]. Instead of these frequency-domain measurements, we perform pump-probe measurements on a staircase-structured hBN to vary the phonon travel distance within a single t-hBN/2L-WSe$_2$/b-hBN/FLG heterostructure (Device 2), where the bottom hBN (b-hBN) has different thicknesses in different regions (see Fig. S1a). Figure 2a shows the schematic illustration of the heterostructure and Fig. 2b shows the measured phonon oscillations. The b-hBN thicknesses ($d_{\text{b-hBN}} = 13.9, 18.6$, and $29.0$ nm) were measured with an atomic force microscope (Fig. S1b-e) and were further confirmed by optical reflection contrast measurements. Using these b-hBN thicknesses along with the measured arrival times of $\Delta t = 4.32, 5.27$, and $7.75$ ps, we obtain an out-of-plane sound speed $v_s = 3500 \pm 130$ m/s for hBN. It yields a homolayer force constant of $K_h = \rho^2 v_s^2/\mu = (8.5 \pm 0.6) \times 10^{19}$ N/m$^3$, where $\rho = 2.298 \times 10^3$ kg/m$^3$ is the 3D density and $\mu = 7.602 \times 10^{-7}$ kg/m$^2$ is the 2D density of hBN. The measured $K_h$ value is consistent with that measured with frequency-domain techniques and theoretical calculations[28–35] (see Table S2).

As an example of THz phononic metamaterial, we constructed a phononic cavity composed of a WSe$_2$/hBN/FLG heterostructure (Device 3, $d_{\text{hBN}} = 19.6$ nm) on an etched substrate, as shown in Fig. 3a,b. The suspended structure allows us to avoid any transmission losses through the substrate. Figure 3c shows the phonon oscillations measured at 77 K. As the phonons travel back and forth between the two ends, the multiple arrivals of phonon pulses are detected by the WSe$_2$ sensing layer. We observe that low-frequency phonons travel faster than higher-frequency phonons due to the phonon dispersion in hBN (see Extended Data Fig. 7b for a longer trace).

By Fourier transforming the pump-probe trace, the power spectrum of the phononic cavity is obtained and shown in Fig. 3d. A THz frequency comb with more than 20 overtone peaks is observed. The extracted line widths ($\delta f$, FWHM) are ~4.1 GHz for all peaks in the range 0.4-2.2 THz (Fig. 3e). The weak frequency dependence indicates the breakdown of Herring processes at temperatures much below the Debye temperature, similar to the saturation of phonon attenuation above 0.7 THz observed in GaAs superlattices[36]. From the extracted line widths, the quality factor ($Q = f/\delta f$) is calculated and plotted as a function of frequency in Fig. 3f. It reaches a very high value of $Q = 500$ at 2 THz, which is unprecedented in the THz regime.

Moreover, the value of $f \times Q = 10^{15}$ Hz at 2 THz (shown in Fig. 3g) far exceeds the value required for ground-state cooling to a phononic quantum state[37] and is the highest reported to date for 2D material systems. A temperature-dependent measurement (Extended Data Fig. 4) also shows that the phonon lifetimes slightly decrease at room temperature, possibly due to additional anharmonic phonon couplings. In some devices, additional phonon loss contribution from surface roughness can be observed. For example, Device 1 shows surface-roughness-induced phonon loss (Extended Data Fig. 5). Using a simple surface roughness model[38], we can estimate a root-mean-square roughness of 0.91 Å for this device.

The free spectral range ($\Delta f$) for each peak, plotted in Fig. 3h, is mostly determined by the total cavity length. However, the frequency-dependent values of $\Delta f$ give information about the phonon dispersion of the heterostructure, which can be simulated using a one-dimensional (1D) nanomechanical model (thick orange line in Fig. 3h; see Methods). The maximum slope (i.e., inflection point) of the dispersion is reached at 0.6 THz and the group velocity decreases at higher frequencies. The decrease of $\Delta f$ at lower frequencies originates from the heavy mass of the WSe$_2$ layer, which results in a reflection phase shift and a change in the cavity resonance condition.

## THz phononic spectroscopy

We further manipulate coherent THz phonons by embedding a monolayer WSe$_2$ in hBN. Figure 4a,b shows the optical microscope image and schematic illustration of our heterostructure composed of t-hBN/WSe$_2$/b-hBN/FLG (Device 4). We chose a top hBN thickness of $d_{\text{t-hBN}} = 83.9$ nm and a bottom hBN thickness of $d_{\text{b-hBN}} = 63.3$ nm so that the reflected and transmitted phonon pulses (2nd and 3rd arrivals, respectively, as marked in Fig. 4b) are separated in time when they arrive at the WSe$_2$ layer. The suspended structure ensures that the reflection at the two end surfaces of the heterostructure is maximized, which allows us to analyze the phonon pulses without any transmission losses through the substrate.

The measured phonon amplitudes $A(t)$ at 300 K are shown with a gray line in Fig. 4d. We used a 10 ps time window for each incident (blue), reflected (orange), and transmitted (green) phonon pulse to analyze their Fourier power spectra $P(f)$, which are plotted in Fig. 4e. The power reflection spectrum is calculated by the ratio between the reflected and incident spectra, $\tilde{R}(f) \equiv P_r(f)/P_i(f)$, and the power transmission spectrum is calculated by the ratio between the transmitted and incident spectra, $\tilde{T}(f) \equiv P_t(f)/P_i(f)$, which are plotted in the left panel of Fig. 4f.

Surprisingly, we found that the reflected phonon power is much higher than the transmitted power, indicating that a single layer of WSe$_2$ can block most phonon transmission at THz frequencies. Note that $\tilde{R}(f)$ and $\tilde{T}(f)$ include different amounts of losses during the propagation over different path lengths. To quantify the reflection and transmission coefficients, we need to account for the intrinsic phonon decay during the propagation and the reflection loss at the end surfaces. This can be achieved using a phenomenological 1D nanomechanical model, as we describe below.

## 1D nanomechanical model

We modeled the observed pump-probe signal and corresponding atomic motion using a 1D chain of masses and springs. Figure 4c shows a representative mass-spring chain, where each layer is modeled as a ball of mass $m$, and the interlayer coupling is modeled with a spring with force constant $K$. We also used a damping element parallel to each spring (Kelvin-Voigt model) to phenomenologically model the decay at interfaces[39,40]. The force constants for homolayer coupling in bulk crystals of graphite ($K_g$), hBN ($K_h$), and WSe$_2$ ($K_w$) are relatively well characterized[28,29,41–45]. However, the force constants of heterolayer coupling ($K_{gh}$ for the graphene-hBN interface and $K_{hw}$ for the hBN-WSe$_2$ interface) are largely unknown and, more importantly, depend on specific configurations of the interface such as the twist angle, defects, inhomogeneity, and roughness[38,46–48]. Therefore, it is crucial to accurately determine the heterolayer coupling parameters for each sample.

By solving the equations of motion (details in Methods), the atomic displacement of each layer can be simulated. To reproduce the observed exciton peak shift, we used the fact that the WSe$_2$ exciton peak energy depends linearly on the hBN-WSe$_2$ gap change[26] (valid in the small displacement regime). The exciton energy shift can be expressed as $\Delta E \propto (x_n - x_{n-1}) + (x_{n+1} - x_n) = x_{n+1} - x_{n-1}$, where $x_n$ is the displacement of the WSe$_2$ layer and $x_{n\pm1}$ are the displacements of the hBN layers adjacent to the WSe$_2$ layer (Fig. 4c).

The simulated exciton peak shift, plotted with a brown line in Fig. 4d, shows excellent agreement with the measured phonon oscillation (gray line). Based on this model, we corrected the propagation losses caused by the different path lengths of reflected and transmitted phonons. In the right panel of Fig. 4f, we show the intrinsic reflection and transmission spectra, $R(f)$ and $T(f)$, at the monolayer WSe$_2$. We note that $R(f) + T(f)$ is smaller than unity, indicating additional interfacial phonon energy loss. This might be due to nonideal heterointerfaces, such as the steric oxygen impurity species protruding out of the plane of the WSe$_2$ layer[48]. Further studies will be needed to understand the loss mechanism of THz phonons at the heterolayer interfaces.

For the simulation, we used homolayer force constants $K_\text{h} = 8.5 \times 10^{19}$ N/m$^3$ and $K_\text{g} = 11.8 \times 10^{19}$ N/m$^3$ obtained from the staircase-structure measurement (Fig. 2) and other Raman measurements[44,45]. We can also obtain important but unknown parameters of heterolayer force constants: $K_\text{gh} = (8.1 \pm 0.4) \times 10^{19}$ N/m$^3$ for the graphene-hBN interface and $K_\text{hw} = (4.0 \pm 0.3) \times 10^{19}$ N/m$^3$ for the hBN-WSe$_2$ interface. The heterolayer force constant values are also reproducible in multiple devices (see Extended Data Fig. 6 and Section 5 of Supplementary Information).

We performed atomistic simulations to obtain the measured theoretical interlayer force constants (see Section 6 of Supplementary Information). To account for the van der Waals interactions, we adopted rVV10 nonlocal density functional. The calculated homolayer force constants are $K_\text{g} = 11.3 \times 10^{19}$ N/m$^3$ and $K_\text{h} = 10.2 \times 10^{19}$ N/m$^3$, which are comparable to the experimental values. However, the *ab initio* calculation predicts the heterolayer force constants to be $K_\text{gh} = 11.1 \times 10^{19}$ N/m$^3$ and $K_\text{hw} = 8.4 \times 10^{19}$ N/m$^3$, higher than the experimentally measured values.

The biggest discrepancy between the experiment and theory is the value of $K_\text{hw}$, where the experimental value is much smaller than the theory prediction. This difference might arise from the nonideal interface between WSe$_2$ and hBN in real devices, where the increased hBN-WSe$_2$ separation due to oxidation-induced WSe$_2$ defects[48] weakens the interlayer coupling and significantly decreases the $K_\text{hw}$ values. Therefore, it is important to experimentally determine the heterolayer mechanical coupling constants in van der Waals heterostructure devices.

**Conclusion and outlook**

In conclusion, we demonstrate broadband THz phonon generation, detection, and manipulation using the van der Waals heterostructure platform. It opens exciting possibilities for THz phononic engineering, which could enable ultrafast acoustic control, quantum phonon operation, as well as new thermal material design. Many applications can benefit from the ultrashort THz phonon wavelength at ~1 nm. Sonic ranging based on THz phonon can enable sensitive detection of embedded interfaces with sub-nanometer spatial resolution. Acousto-optics based on THz phonons will allow acoustic control of EUV and X-ray photons of nm wavelengths for the first time. The large energy bandwidth of THz phonons also allows us to access a large phase space of the phonon Brillouin zone. This is very important for understanding and controlling thermal transport, which is determined by thermal phonons up to 6 THz at room temperature. By designing THz phononic cavities and phononic metamaterials, artificial thermal insulators better than any natural materials might be realizable. Furthermore, the ultrahigh $f \times Q$ value of $10^{15}$ for THz phonons in hBN could enable quantum transduction between THz phonons and visible photons in integrated phononic- and photonic-crystal hBN cavities in the future.

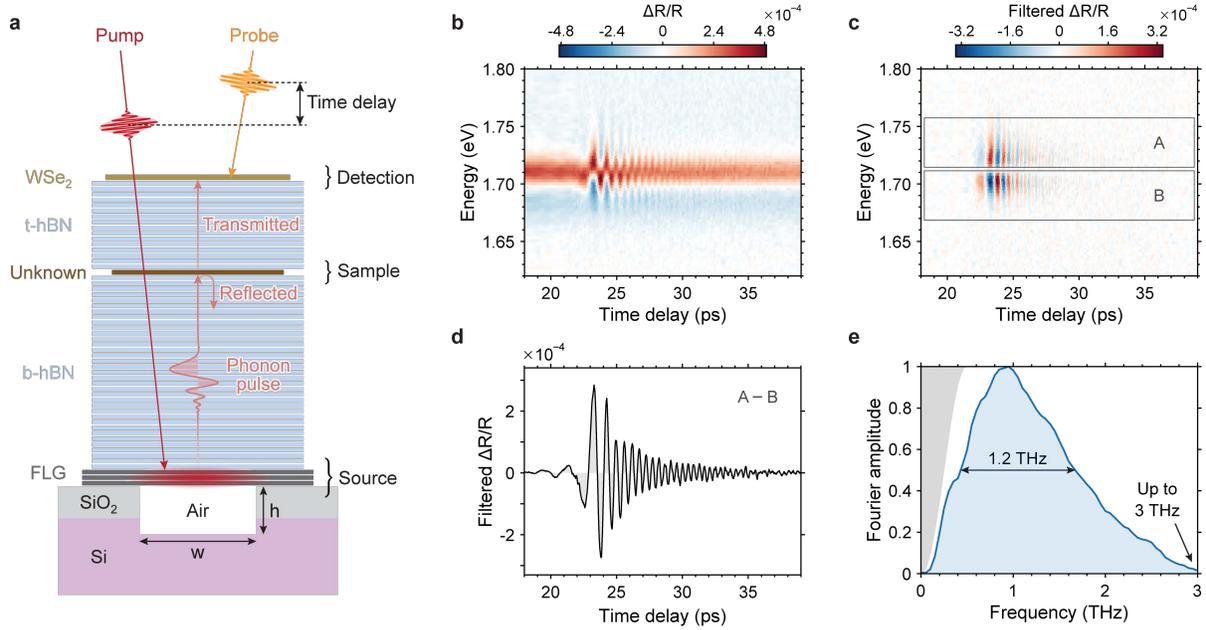

**Fig. 1 | THz phonon spectroscopy with van der Waals heterostructures. a**, Concept of phononic spectroscopy by analogy with optical spectroscopy. A pump pulse ($E_{\text{pump}} = 1.409$ eV) induces an ultrafast expansion of few-layer graphene (FLG), launching an acoustic-phonon pulse that propagates across the hBN spacer. The phonon pulse interacts with an unknown layer and is then detected by a monolayer WSe$_2$. A broadband probe pulse ($E_{\text{probe}} = 1.55\text{-}1.91$ eV) with a time delay $\Delta t$ monitors the exciton peak shift induced by the phonon pulse. The silicon substrate is etched so that the well depth ($h$) is 120-160 nm and the width ($w$) is 20-25 μm. **b**, Using a WSe$_2$/hBN/3LG heterostructure (Device 1, $d_{\text{hBN}} = 80.2$ nm), the transient reflection ($\Delta R/R$) spectrum of the WSe$_2$ exciton peak at 77 K shows phonon-induced exciton oscillations. **c**, The DC component is filtered out to show only the oscillating components. **d**, Difference between the high-energy (A) and low-energy (B) sides of the exciton peak. **e**, Fourier amplitude spectrum of the phonon pulse with FWHM of 1.2 THz and spectral content up to 3 THz. Gray shaded region shows the DC filter spectrum.

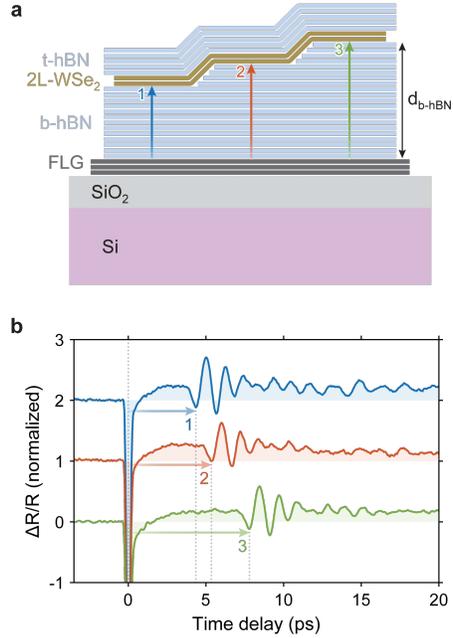

**Fig. 2 | Determination of the phonon propagation speed in hBN. a**, Schematic illustration of a staircase heterostructure, t-hBN/2L-WSe$_2$/b-hBN/FLG (Device 2), with three regions of b-hBN thicknesses: $d_\text{b-hBN}$ = 16.2, 20.3, and 28.6 nm. **b**, $\Delta R/R$ measured in each region shows varying time delays in phonon oscillations: $\Delta t$ = 4.3, 5.3, and 7.7 ps, respectively. The measured $\Delta R/R$ values were normalized to the optical AC Stark shift signal at $t = 0$, and then scaled and shifted for clarity.

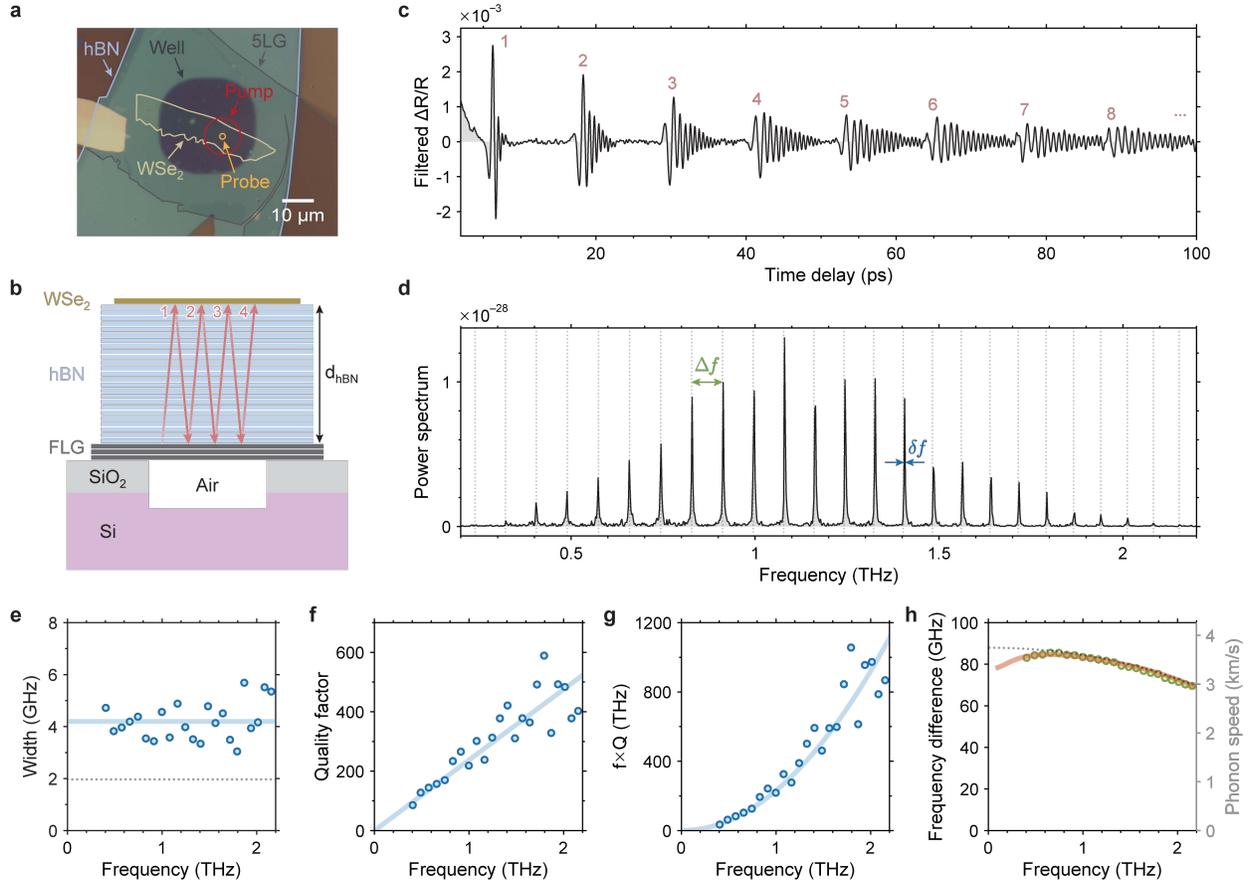

**Fig. 3 | THz phononic cavity and frequency comb. a,b**, A microscope image and schematic illustration of a THz phononic cavity structure, WSe$_2$/hBN/5LG (Device 3, $d_{hBN}$ = 19.6 nm). **c**, Phonon oscillations measured at 77 K. **d**, The power spectrum of phonon oscillations from 3 ps to 453 ps shows frequency-comb-like peaks. **e**, The line width ($\delta f$), defined as the FWHM of each peak, is plotted in blue circles. The width given by the 450 ps rectangular time window (i.e., FWHM of the sinc function) is plotted with a dotted gray line. **f**, Quality factor ($Q = f/\delta f$) of each peak (blue circles) and a linear fit (thick blue line). **g**, The $f \times Q$ product of each peak (blue circles) and a quadratic fit (thick blue line). **h**, Frequency-dependent free spectral range ($\Delta f$) shows that the hBN phonon dispersion has an inflection point at 0.6 THz. The $\Delta f$ values simulated using the 1D mass-spring model are plotted with a thick orange line. The phonon speed calculated for a bulk hBN is plotted in a gray dotted line (right axis).

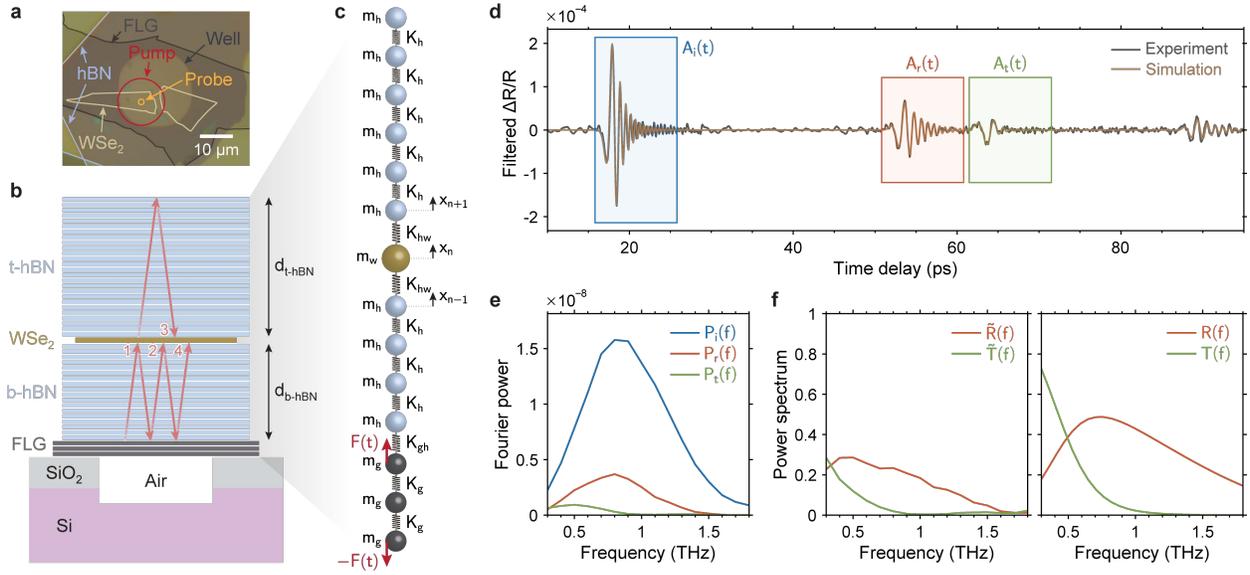

**Fig. 4 | THz reflection/transmission spectra and 1D mass-spring model simulation. a,b**, A microscope image and schematic illustration of a t-hBN/WSe$_2$/b-hBN/3LG heterostructure (Device 4, $d_{\text{t-hBN}} = 83.9$ nm and $d_{\text{b-hBN}} = 63.3$ nm) designed for the measurement of THz reflection and transmission spectra of a monolayer WSe$_2$ embedded in hBN. **c**, Corresponding 1D mass-spring model. Red arrows denote the initial laser-induced forces. **d**, Time-domain phonon oscillations measured at 300 K. The gray line shows the measured pump-probe trace and the brown line shows the results of the 1D mass-spring simulation. **e**, Fourier power of the incident (blue), reflected (orange), and transmitted (green) phonon pulses (the time windows are shown in blue, orange, and green rectangles in **d**). Dashed lines are the results of the simulation. **f**, The left panel shows uncorrected reflection (orange, $\tilde{R}$) and transmission (green, $\tilde{T}$) spectra at the monolayer WSe$_2$ embedded in hBN. Dashed lines are the results of the simulation. The right panel shows corresponding THz reflection/transmission spectra ($R$ and $T$) corrected for the propagation loss, which reveals that a finite energy loss is present at the heterolayer interface.

## Methods

Device fabrication and characterization

The devices were fabricated using a dry-transfer technique based on polycarbonate (PC). We first exfoliated WSe$_2$, hBN, and FLG from bulk crystals onto SiO$_2$/Si substrates and then vertically assembled them into the desired heterostructure using a PC stamp. We also etched a 90 nm SiO$_2$/Si substrate using 6 sccm O$_2$ and 80 sccm CHF$_3$ at a power of 250 W for a few minutes. A well depth of 120-160 nm (to maximize the local field at the WSe$_2$ layer depending on the specific stacking order and thicknesses) and a lateral diameter of 20-25 μm were chosen. Finally, the whole heterostructure was released onto the etched SiO$_2$/Si substrate to achieve the suspended devices. The thickness of each layer is estimated using optical reflection contrast spectra and calibrated atomic force microscope scans. Twist angles between layers are not controlled.

Pump-probe spectroscopy

Yb:KGW femtosecond laser (Light Conversion CARBIDE, wavelength λ = 1030 nm, repetition rate $f_{\text{rep}}$ = 75.6 kHz) was used as the "fundamental" beam for an optical parametric amplifier (OPA, Light Conversion ORPHEUS). The OPA signal energy was tuned to $E_{\text{pump}}$ = 1.409 eV ($\lambda_{\text{pump}}$ = 880 nm) and used as a "pump" beam so that it is only absorbed by FLG but not by any of the TMD layers. The leftover fundamental beam was focused onto a 3-mm-thick sapphire crystal to generate a supercontinuum, which was filtered to let through $E_{\text{probe}}$ = 1.55-1.91 eV and used as a "probe" beam. The pulse durations (FWHM of the intensity envelope) of both pump and probe pulses were initially 200 fs but elongated to 250 fs after passing through multiple optical components including an objective lens (20×, NA 0.45). Both beams were vertically polarized for pump-probe measurements. Circular-polarization-resolved measurements showed that the co- and cross-polarization signals ($\sigma^+\sigma^+$ and $\sigma^+\sigma^-$) differed only by the optical AC Stark shift signal[49,50] that shows up when the pump and probe pulses overlap in time ($\Delta t = 0$). The FWHM of the pump beam's spatial profile was 10-15 μm (varied for different experiments), while that of the probe beam was 1.6 μm. The reflected probe beam was dispersed by a transmissive grating (300 gr/mm), focused by a lens (5 cm focal length), and detected by a linear camera (Coptonix, S11639-01). The pump beam was chopped at 707 Hz and the pump-probe signal $\Delta R/R$ was averaged for many thousands of frames at each time delay. Group delay dispersion induced by optical components is post-corrected during the analysis. The exact values of these parameters (pump beam size, integration time, etc.) were varied for each measurement to optimize the signal.

1D nanomechanical (mass-spring chain) model

For the 1D mass-spring model shown in Fig. 4c (t-hBN/WSe$_2$/b-hBN/3LG), the equation of motion for the very bottom graphene layer is

$$m_1\ddot{x}_1 = K_g(x_2 - x_1) + \gamma_g(\dot{x}_2 - \dot{x}_1) - F(t), \tag{2}$$

where $x_i$ is the displacement of $i$th layer, $F(t) = F_0 e^{-t^2/2\sigma^2}$ is the Gaussian function describing the laser-induced thermal expansion of FLG (FWHM of 400 fs, estimated from Extended Data Fig. 1), and $\gamma_g$ is the viscoelastic damping coefficient for the spring connecting the first and second graphene layers. The equations of motion for the other two graphene layers are

$$m_2\ddot{x}_2 = -K_g(x_2 - x_1) + K_g(x_3 - x_2) - \gamma_g(\dot{x}_2 - \dot{x}_1) + \gamma_g(\dot{x}_3 - \dot{x}_2) \tag{3}$$

$$m_3\ddot{x}_3 = -K_g(x_3 - x_2) + K_{gh}(x_4 - x_3) - \gamma_g(\dot{x}_3 - \dot{x}_2) + \gamma_{gh}(\dot{x}_4 - \dot{x}_3) + F(t). \tag{4}$$

Similarly, equations of motion for all layers can be written down and solved numerically using the conditions identical to the experiment (time steps, time window, etc.). The exciton energy shift is estimated as $\Delta E \propto (x_n - x_{n-1}) + (x_{n+1} - x_n) = x_{n+1} - x_{n-1}$ for embedded WSe$_2$ structures (e.g., hBN/WSe$_2$/hBN/FLG) and as $\Delta E \propto x_n - x_{n-1}$ for end WSe$_2$ structures (e.g., WSe$_2$/hBN/FLG), where $x_n$ is the displacement of the WSe$_2$ layer.

To reproduce the measured time trace and Fourier spectra in Fig. 4d-f, we used 3 graphene layers, 180 bottom hBN layers, 1 WSe$_2$ layer, 238 top hBN layers, and the phenomenological damping rates of $\gamma_g = \gamma_h = (30 \text{ ps})^{-1}$, $\gamma_{gh} = (4 \text{ ps})^{-1}$, and $\gamma_{hw} = (2 \text{ ps})^{-1}$ to account for the interface losses due to imperfect stacking and defects introduced during fabrication steps. Additional damping at the top hBN layer of $\gamma = (0.8 \text{ ps})^{-1}$ was added to reproduce the transmission loss through contamination such as polymer residue. The force constants used for the simulation are given in the main text.

On the other hand, to solve for the eigenvalues of the heterostructure phonon modes, we used the same equations of motion but without the time-varying force term $F(t)$. By replacing $x_n$ with $A_n e^{i\omega t}$ in the equations of motion, the coupled differential equations can be modified to an eigenvalue problem with a $N \times N$ matrix, where $N$ is the total number of layers ($N = 240$ for Fig. 1). This procedure gives the phonon eigenmode frequencies of the entire heterostructure.

Atomic simulation of force constants

To calculate the interlayer force constants of heterointerfaces, we performed *ab initio* density-functional theory (DFT) and density-functional perturbation theory (DFPT) calculations[51]. To simulate heterointerfaces, we constructed homo- and hetero-bilayer structures of graphene, hBN, hBN-graphene, and hBN-WSe$_2$. For graphene and hBN homo-bilayer, we adopted natural bilayer stacking configurations: AB (or Bernal) and AA' stackings for graphene and hBN bilayers, respectively. For the hBN-graphene hetero-bilayer structure we construct a 21.8°-twisted bilayer structure to mimic the lack of alignment between the layers in the experiment (Extended Data Fig. 8a). For the hBN-WSe$_2$ hetero-bilayer, we constructed a bilayer structure with a 4×4 supercell of hBN and 3×3 supercell of WSe$_2$ to account for their different lattice constants (Extended Data Fig. 8b). We relaxed the constructed bilayer structures by minimizing the DFT total energy. Subsequently, we calculated the layer-breathing mode frequencies and corresponding interlayer force constants by performing DFPT calculations.

The DFT and DFPT calculations were performed using the plane-wave Quantum ESPRESSO package[52] and using optimized norm-conserving Vanderbilt pseudopotentials[53,54]. To account for the van der Waals interactions, we adopted rVV10 nonlocal density functional[55]. Additionally, we included the calculation results using local density approximation (LDA) functional[56]. Previous studies have demonstrated that LDA functional provides more reasonable estimations of the layer-breathing-mode frequencies for multilayer graphene[57-61] and hBN[28,35,62] structures. For the semimetallic graphene and graphene-hBN calculations, we applied 0.02 Ry of Gaussian smearing for the electronic states. We ensured the convergence of the breathing mode frequency for each structure within 5 cm$^{-1}$ with respect to the *k*-point sampling and planewave energy cutoff. The DFT parameters and calculated interlayer force constants are summarized in Tables S3 and S4 of Supplementary Information.

## Acknowledgements

The pump-probe spectroscopy, data analysis, as well as DFT and DFPT calculations are supported by the U.S. Department of Energy, Office of Science, Basic Energy Sciences, Materials Sciences and Engineering Division under Contract No. DE-AC02-05CH11231 within the Nanomachine Program. The monolayer exfoliation and heterostructure stacking are supported by the U.S. Department of Energy, Office of Science, Basic Energy Sciences, Materials Sciences and Engineering Division under Contract No. DE-AC02-05CH11231 within the van der Waals heterostructure program (KCWF16). The suspended device fabrication is supported by the U.S. Army Research Office under MURI award W911NF-17-1-0312. S.C. acknowledges support from the Kavli ENSI Heising-Simons Junior Fellowship. W.K., M.H.N., and S.G.L. acknowledge the Texas Advanced Computing Center (TACC) at the University of Texas at Austin for providing high-performance computing resources. This research also used resources of the National Energy Research Scientific Computing Center (NERSC), a U.S. Department of Energy Office of Science User Facility located at Lawrence Berkeley National Laboratory, operated under Contract No. DE-AC02-05CH11231. K.W. and T.T. acknowledge support from the JSPS KAKENHI (Grant Numbers 21H05233 and 23H02052) and World Premier International Research Center Initiative (WPI), MEXT, Japan.

## Author contributions

Y.Y. and F.W. conceived the project. F.W., M.F.C., and S.G.L. supervised the project. Y.Y. and C.U. performed pump-probe measurements. Z.L., Y.Y., R.Q., W.Z., S.C., and Q.F. fabricated and characterized the samples. Y.Y. and Z.L. performed 1D mass-spring simulations. W.K. and M.H.N. performed DFT-based simulations. K.W. and T.T. grew hBN crystals. All the authors discussed the results and contributed to the manuscript.

## Competing interests

The authors declare no competing interests.

## Data availability

The data that support the findings of this study are available from the corresponding authors upon request.

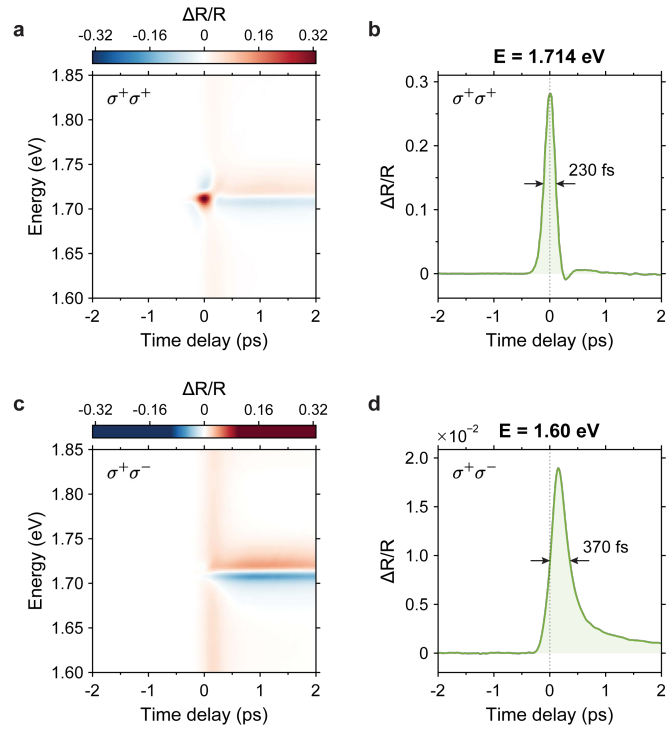

**Extended Data Fig. 1 | Fast hot-carrier relaxation in FLG. a**, A near-infrared (NIR) pump-probe spectrum measured in an hBN/WSe$_2$/hBN/FLG heterostructure at 77 K. The pump pulse was at $E_{\text{pump}} = 1.409$ eV and right-circularly polarized ($\sigma^+$). A broadband probe pulse ($E_{\text{probe}} = 1.55\text{-}1.91$ eV) with the same polarization ($\sigma^+$) measures the pump-induced change of the heterostructure reflectance. **b**, A horizontal line cut at $E = 1.714$ eV shows the optical AC Stark shift at time zero ($\Delta t = 0$), which follows the pump pulse intensity envelope (FWHM of 230 fs). **c**, A pump-probe spectrum measured with a left-circularly polarized ($\sigma^-$) probe pulse, where the optical AC Stark shift signal is absent. The color scale is saturated at the ±30% of the maximum level in **a**. **d**, A horizontal line cut at $E = 1.60$ eV shows that the hot-carrier-induced broadband reflectance change of FLG appears and disappears in ~370 fs (FWHM).

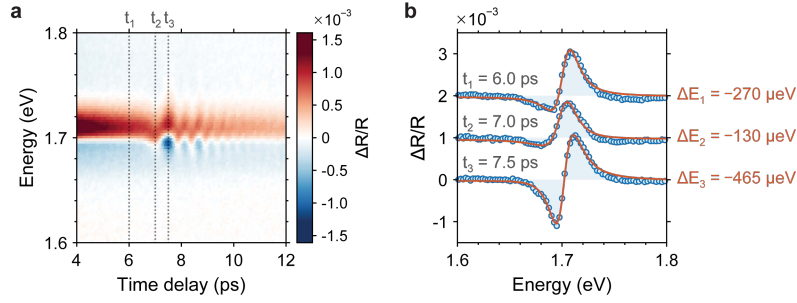

**Extended Data Fig. 2 | Phonon-induced exciton energy shift. a**, Pump-probe spectra ($\Delta R/R$) measured in an hBN/WSe$_2$/hBN/3LG heterostructure ($d_{\text{t-hBN}} = 41.0$ nm and $d_{\text{b-hBN}} = 24.6$ nm) at 77 K. The pump fluence was 250 µJ/cm$^2$. **b**, Vertical line cuts at $t_1 = 6.0$ ps, $t_2 = 7.0$ ps, and $t_3 = 7.5$ ps are shown in blue circles. The data at $t_1$ and $t_2$ are vertically shifted for clarity. By performing transfer-matrix calculations including pump-induced exciton energy shift ($\Delta E = E_{\text{pump on}} - E_{\text{pump off}}$), line broadening ($\Delta \Gamma = \Gamma_{\text{pump on}} - \Gamma_{\text{pump off}}$), and oscillator strength change ($\Delta A = A_{\text{pump on}} - A_{\text{pump off}}$) as fitting parameters, the fit results (orange lines) and corresponding energy shift values are shown. The exciton energy shift between $t_2$ and $t_3$ is $|\Delta E_3 - \Delta E_2| = 335$ µeV.

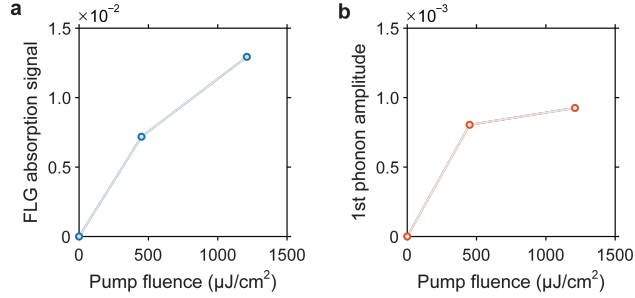

**Extended Data Fig. 3 | Saturation of phonon amplitude at high pump fluence. a**, The pump-induced FLG absorption signal of Device 3 at $\Delta t = 0$ (similar to that shown in Extended Data Fig. 1d) at pump fluence of $450~\mu J/cm^2$ and $1210~\mu J/cm^2$ show a roughly linear increase of the FLG absorption as a function of pump fluence (weak saturation). **b**, The peak-to-peak amplitude of the first phonon pulse at $\Delta t = 6.2$ ps at pump fluence of $450~\mu J/cm^2$ and $1210~\mu J/cm^2$ show strong saturation at the high pump fluence. The sublinear dependence in **a** implies that not all pump power is absorbed by FLG, and the stronger sublinear dependence in **b** implies that not all absorbed power is converted to phonon amplitudes at high pump fluence.

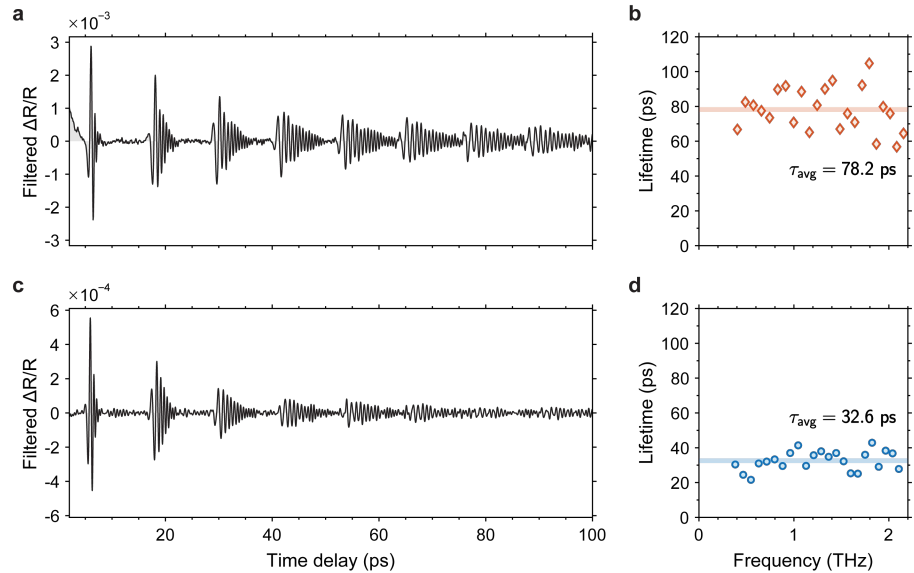

**Extended Data Fig. 4 | Temperature-dependent phonon oscillation decay.** Phonon oscillations of Device 3 at 77 K (**a**, **b**) and at 300 K (**c**, **d**) with similar pump fluences (300-400 μJ/cm$^2$, lower than the saturation regime). Hann filters are used to remove the DC component. Note that the absolute pump fluences cannot be directly compared due to the different local field strengths at the FLG location at these two temperatures. The phonon amplitude lifetimes are calculated by the procedure described in Section 4 of Supplementary Information.

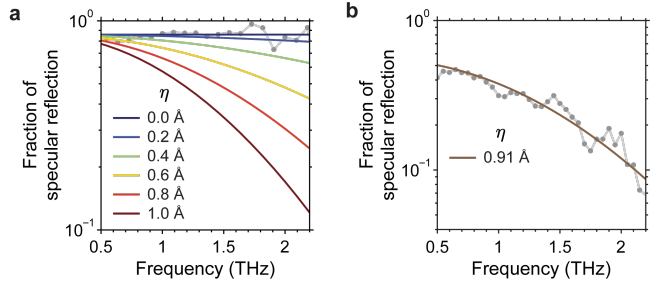

**Extended Data Fig. 5 | Frequency-dependent phonon loss due to surface roughness. a**, For Device 3 (WSe$_2$/hBN/5LG, $d_{\text{hBN}} = 19.6$ nm), the fraction of specular reflection after a round trip, $p$, is plotted as a function of phonon frequency. The gray-filled circles are obtained from the Fourier analysis of phonon pulses. This behavior can be modeled by including both the intrinsic hBN phonon loss and an additional surface-roughness-induced loss. Based on the model described by J. M. Ziman[38], the fraction of specular reflection can be expressed as $p(\lambda) = \exp(-16\pi^3\eta^2/\lambda^2)$, where $\lambda$ is the phonon wavelength and $\eta$ is the root-mean-square deviation of the interfacial height from a reference plane. The colored lines show calculated $p(\lambda)$ with various $\eta$ values. It indicates that the surface roughness is below 0.3 Å in Device 3. **b**, For Device 1 (WSe$_2$/hBN/3LG, $d_{\text{hBN}} = 80.2$ nm), the fraction of specular reflection after a round trip is plotted in gray-filled circles. The brown line shows the best-fit result with a roughness value of $\eta = 0.91$ Å.

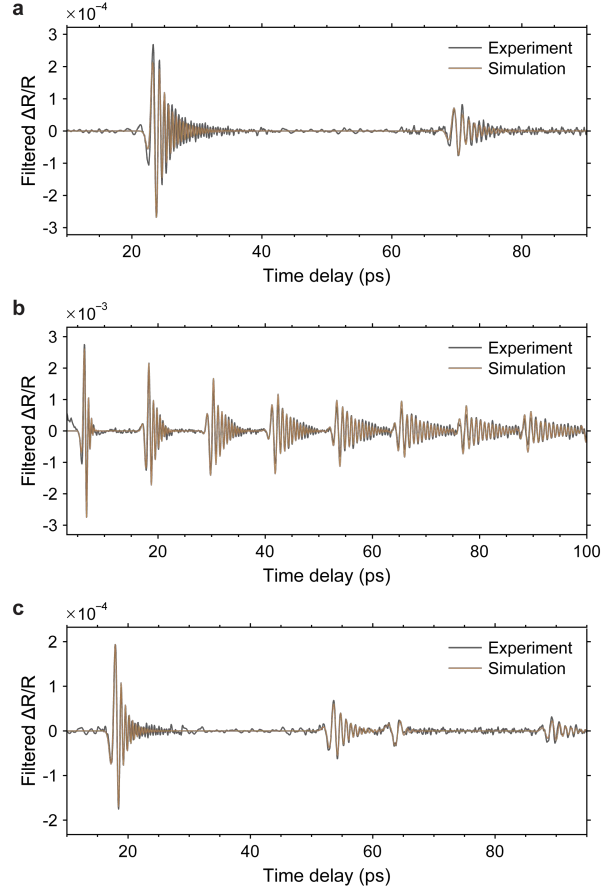

**Extended Data Fig. 6 | 1D nanomechanical simulation of Devices 1, 3, and 4. a**, Phonon oscillation data (gray line) and the simulation result (brown line) using $K_{gh} = 7.3 \times 10^{19}$ N/m³ and $K_{hw} = 3.8 \times 10^{19}$ N/m³ for Device 1. **b**, Phonon oscillation data (gray line) and the simulation result (brown line) using $K_{gh} = 8.8 \times 10^{19}$ N/m³ and $K_{hw} = 4.6 \times 10^{19}$ N/m³ for Device 3. **c**, Phonon oscillation data (gray line) and the simulation result (brown line) using $K_{gh} = 8.1 \times 10^{19}$ N/m³ and $K_{hw} = 4.0 \times 10^{19}$ N/m³ for Device 4. On average, the heterolayer force constant values are $K_{gh} = (8.1 \pm 0.4) \times 10^{19}$ N/m³ and $K_{hw} = (4.0 \pm 0.3) \times 10^{19}$ N/m³ across the three devices.

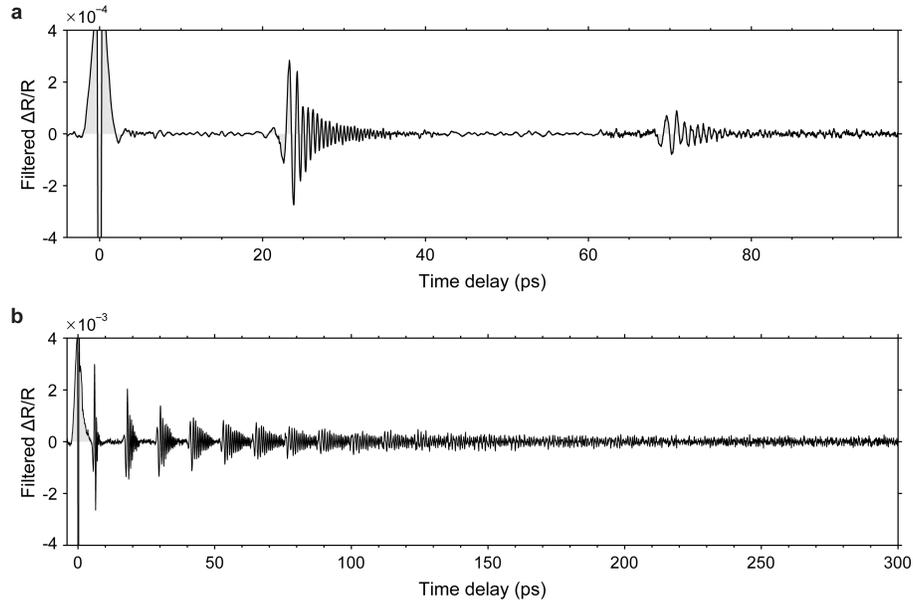

**Extended Data Fig. 7 | Longer time traces of the data shown in the main text. a**, Phonon oscillation data identical to that in Fig. 1d but with a longer time window (Device 1, WSe$_2$/hBN/3LG, $d_{hBN}$ = 80.2 nm, at 77 K). **b**, Phonon oscillation data identical to that in Fig. 3c (Device 3, WSe$_2$/hBN/5LG, $d_{hBN}$ = 19.6 nm, at 77 K) is shown up to 300 ps.

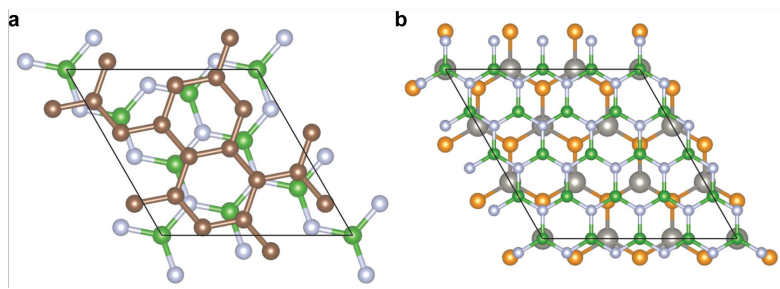

**Extended Data Fig. 8 | Simulated atomic structure of heterointerface. a**, Simulated atomic structure of the graphene-hBN heterobilayer. **b**, Simulated atomic structure of the hBN-WSe$_2$ heterobilayer.

# Supplementary Information for
# "Terahertz phonon engineering with van der Waals heterostructures"

Yoseob Yoon[1,2,3,*], Zheyu Lu[2,4], Can Uzundal[2,5], Ruishi Qi[1,2], Wenyu Zhao[1], Sudi Chen[1,6], Qixin Feng[1,2], Woochang Kim[1,2], Mit H. Naik[1,2], Kenji Watanabe[7], Takashi Taniguchi[8], Steven G. Louie[1,2], Michael F. Crommie[1,2,6], and Feng Wang[1,2,6,*]

[1]Department of Physics, University of California, Berkeley, CA, USA

[2]Materials Sciences Division, Lawrence Berkeley National Laboratory, Berkeley, CA, USA

[3]Department of Mechanical and Industrial Engineering, Northeastern University, Boston, MA, USA

[4]Graduate Group in Applied Science and Technology, University of California, Berkeley, CA, USA

[5]Department of Chemistry, University of California, Berkeley, CA, USA

[6]Kavli Energy NanoScience Institute, Berkeley, CA, USA

[7]Research Center for Electronic and Optical Materials, National Institute for Materials Science, Tsukuba, Japan

[8]Research Center for Materials Nanoarchitectonics, National Institute for Materials Science, Tsukuba, Japan

[*]Correspondence to: y.yoon@northeastern.edu; fengwang76@berkeley.edu

# Contents

# 1. Estimation of the out-of-plane sound speed in hBN

The error of the measured sound speed in hBN, $v_s = d_{\text{hBN}}/\Delta t$, comes from the measurements of hBN thickness ($d_{\text{hBN}}$) and time delay ($\Delta t$). First, we estimated the thickness of the staircase b-hBN (shown in Fig. 2a of the main text) using an atomic force microscope (Bruker Innova, calibrated using VGRP-18M with 177 nm depth and APCS-0099 with 107 nm depth). We used a contact mode to scan 4 μm×4 μm and 9 μm×9 μm regions (denoted as ① and ② in Fig. S1a) and gathered the line cuts at each $y$ value. The line-cut height profiles, $z(x, y)$, are fitted the data with a function

$$z(x,y) = \frac{c_1}{2}\text{erfc}\left(\frac{x - c_2}{c_3}\right) + c_4 x^3 + c_5 x^2 + c_6 x + c_7, \tag{S1}$$

where erfc is the complementary error function and the hBN thickness is parameterized by the coefficient $c_1$. Representative line cuts and fits are shown in Fig. S1c,d.

From the scan in Region ① (Fig. S1b), the thickness of thick b-hBN is estimated to be $d_{\text{thick}} = 29.0 \pm 1.2$ nm (95% confidence interval). On the other hand, the scan in Region ② shows that the differences between $d_{\text{thick}}$, $d_{\text{medium}}$, and $d_{\text{thin}}$ are $d_{\text{thick}} - d_{\text{medium}} = 4.7 \pm 0.1$ nm and $d_{\text{medium}} - d_{\text{thin}} = 10.4 \pm 0.2$ nm, where the uncertainties are estimated from the 95% confidence interval. From these, the thicknesses of thick, medium, and thin b-hBN are estimated as $d_{\text{thick}} = 29.0 \pm 1.2$ nm, $d_{\text{medium}} = 18.6 \pm 1.2$ nm, and $d_{\text{thin}} = 13.9 \pm 1.2$ nm, where the dominating source of error was from the scan in Region ①.

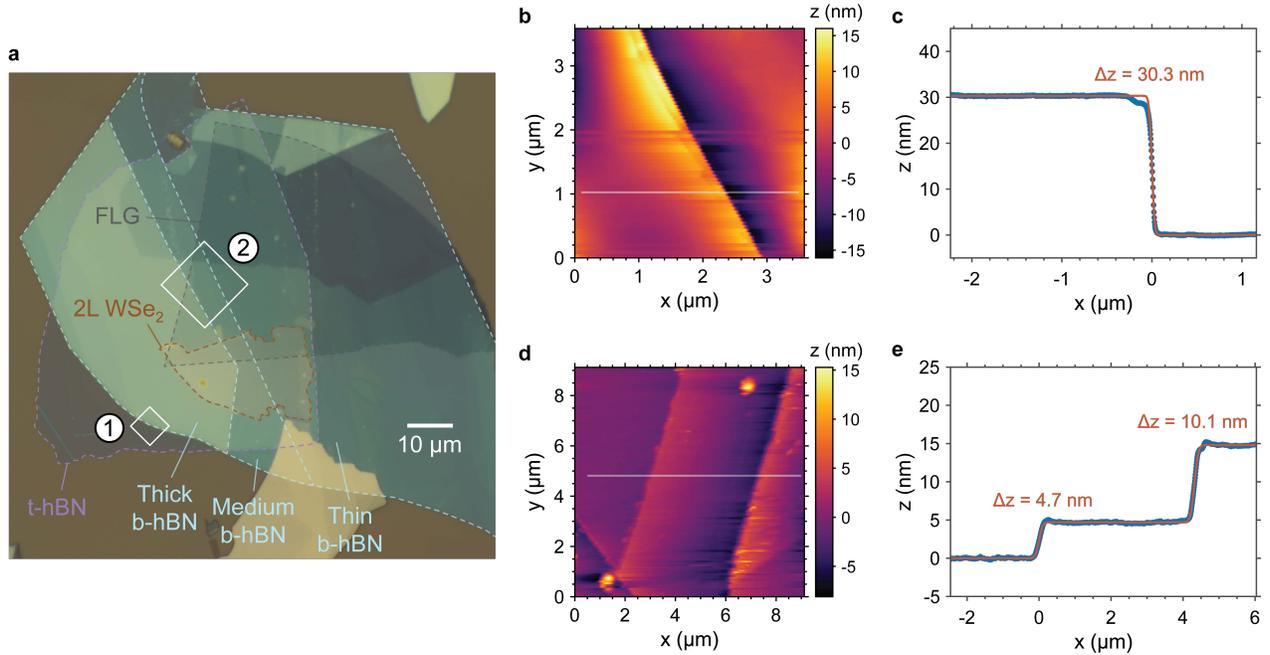

**Fig. S1 | Atomic Force Microscope (AFM) characterization of hBN thicknesses. a**, Optical microscope image of Device 2 used for Fig. 2 of the main text. **b**, An AFM scan in Region ①. A representative line cut (marked as a white horizontal line) is shown in **c**. The background-subtracted data is shown with blue circles, and the best fit to the function Eq. (S1) is shown with an orange line. **d**, An AFM scan in Region ②. A representative line cut (marked as a white horizontal line) is shown in **e**. The background-subtracted data is shown with blue circles, and the best fit to the function Eq. (S1) is shown with an orange line.

On the other hand, the error in $\Delta t$ is negligible. The delay stage has an accuracy of $\pm 2.25$ μm and unidirectional repeatability of $\pm 1$ μm, which correspond to $\pm 30$ fs and $\pm 13.3$ fs, respectively, in the double-pass geometry used in all the measurements. The time step size was set to $\Delta t_{\text{step}} = 50\text{-}100$ fs. The initial phonon pulse duration is about 370 fs (shown in Extended Data Fig. 1d), but the determination of the arrival time in Fig. 2b was not limited to 370 fs but limited by the ability to find the phonon peak position (estimated to be $\sim 50$ fs).

Table S1 summarizes the measured hBN thickness and corresponding phonon propagation speed. The average sound speed in hBN is estimated to be $3500 \pm 130$ m/s.

|  | Thickness (nm) | Arrival time (ps) | Sound speed (m/s) |
|---|---|---|---|
| Thick region | 29.0 ± 1.2 | 7.75 ± 0.05 | 3740 ± 160 |
| Medium region | 18.6 ± 1.2 | 5.27 ± 0.05 | 3530 ± 230 |
| Thin region | 13.9 ± 1.2 | 4.32 ± 0.05 | 3220 ± 280 |
| Average | - | - | 3500 ± 130 |

**Table S1 | Measured hBN sound speed in three regions of different hBN thicknesses.** The thickness uncertainties are estimated from the 95% confidence interval of a set of AFM line-scan measurements. The arrival time uncertainties are estimated from various sources. The uncertainties are propagated through the calculation of the sound speed $v_{\text{s}} = d_{\text{hBN}}/\Delta t$.

## 2. Measured hBN sound speed and comparison to other reported values

There are many different ways to measure the out-of-plane sound speed in hBN. In Table S2, we summarize the reported values from the literature. The measured values vary between different methods. The dominant uncertainty in the pump-probe measurement was from the hBN thickness determination (as described in Section 1), which can be reduced by careful AFM measurements.

|  | Method | $C_{33}$ (GPa) | $\hbar\omega$ (meV) | $v_{\text{s}}$ (m/s) | $K_{\text{h}}$ ($10^{19}$ N/m$^3$) |
|---|---|---|---|---|---|
| Jiménez-Riobóo et al.[30] | Brillouin light scattering | 24.5 ± 0.2 | - | 3270 ± 10 | 7.41 ± 0.06 |
| Lynch et al.[31] | Compressibility | 32.4 ± 3.0 | - | 3760 ± 170 | 9.79 ± 0.91 |
| Bosak et al.[32] | Inelastic X-ray scattering | 27.0 ± 0.5 | - | 3430 ± 30 | 8.16 ± 0.15 |
| Serrano et al.[28] | Inelastic X-ray scattering | 32.7 ± 6.5 | 15.0 ± 1.5[a] | 3770 ± 380 | 9.87 ± 1.97 |
| Green et al.[33] | Theory | 31.2 | - | 3680 | 9.43 |
| Ohba et al.[34] | Theory | 28.2 | - | 3500 | 8.52 |
| Jiang et al.[35] | Theory | 32.8 | - | 3780 | 9.92 |
| Stenger et al.[29] | Raman scattering | N/A | N/A[b] | N/A | N/A |
| This work (Yoon et al.) | Phonon spectroscopy | 28 ± 2 | - | 3500 ± 130 | 8.5 ± 0.6 |

**Table S2 | Literature values of the hBN sound speed $v_{\text{s}}$ and interlayer force constant $K_{\text{h}}$.** The $C_{33}$ values are reported in the literature, except for Serrano et al.[28] (phonon energy $\hbar\omega$ is reported), Jiang et al.[35] ($v_{\text{s}}$ is reported) and this work ($v_{\text{s}}$ is reported). Other values are deduced by using $v_{\text{s}} = \sqrt{C_{33}/\rho}$, $K = \rho^2 v_{\text{s}}^2/\mu$, and $\omega = \sqrt{4K/\mu}$, where $\rho$ [kg/m$^3$] and $\mu$ [kg/m$^2$] are 3D and 2D densities, respectively. [a]The error ±1.5 meV was estimated from the instrumental energy resolution of 3.0 meV. [b]The out-of-plane B$_{1g}$ phonon mode was not observed in low-frequency Raman measurements.

## 3. Quasi-DC signal filtering procedure

After the strong optical AC Stark shift signal at $t = 0$, there is a slowly varying and long-lived quasi-DC signal at the WSe$_2$ exciton energy at $t > 0$. This quasi-DC signal can be attributed to the free carriers in WSe$_2$ generated by two-photon absorption: $2E_{\text{pump}} = 2.818 \text{ eV} > E_{\text{exciton}}$. These free carriers are instantly generated at $t = 0$ and quickly relax to the exciton state. The emitted optical phonons, relaxed exciton population, and residual hot carriers lead to the long-lived red shift, saturation, and broadening of the exciton peak. This effect gives rise to a slowly increasing DC background that dominates the phonon spectrum. Thus, we process the data by multiplying a Hann window $w(f)$ in the Fourier domain

$$w(f) = \begin{cases} 1 & |f| \geq f_{\text{cut}} \\ 1 - \cos^2\left(\frac{\pi}{2}\frac{f}{f_{\text{cut}}}\right) & |f| < f_{\text{cut}} \end{cases}$$

with a frequency cut-off of $f_{\text{cut}} = 0.4$ or $0.5$ THz, which is shown in Fig. 1e as a gray shaded area. The quasi-DC background can be suppressed. Figure S2 shows the data before and after the processing. Note that the strong optical AC Stark shift signal at $t = 0$ shows positive signals near the $t = 0$ peak, which are artifacts of the filtering. This also appears in Fig. 3c and Extended Data Fig. 5 (and all other data but not shown).

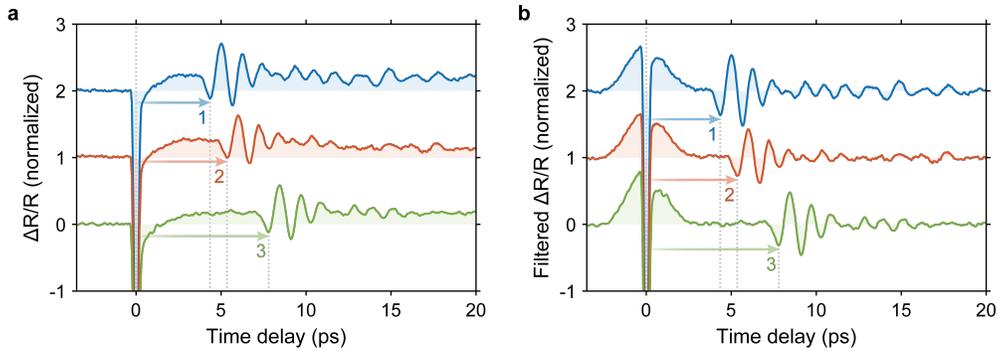

**Fig. S2 | Comparison between raw and filtered $\Delta R/R$ data. a**, Raw $\Delta R/R$ data identical to Fig. 2b. **b**, Filtered $\Delta R/R$ processed in the same way as the rest of the data presented in this work. The Hann window suppresses the quasi-DC background in the time domain.

## 4. Line width, quality factor, and amplitude decay time

The measured $\Delta R/R$ is proportional to the WSe$_2$ exciton energy shift $\Delta E$ (refs. [49,50]), which is also linearly dependent on the phonon amplitude $A(t)$ (ref. [26]). The phonon amplitude decay can be modeled as

$$A(t) = \begin{cases} 0, & t < 0 \\ A_0 e^{i2\pi f_0 t} e^{-\gamma t}, & t \geq 0 \end{cases} \tag{S2}$$

where $A_0$ is the initial phonon amplitude after generation, $f_0$ is the central frequency, and $\gamma = 1/\tau$ is the amplitude decay rate ($\tau$ is the amplitude lifetime). The Fourier transform of $A(t)$ for $t \geq 0$ is

$$\tilde{A}(f) = \int_0^\infty A_0 e^{i2\pi f_0 t} e^{-\gamma t} e^{-i2\pi f t} dt$$

$$= A_0 \int_0^\infty e^{-(i2\pi(f-f_0)+\gamma)t} dt$$

$$= \frac{A_0/2\pi}{i(f-f_0) + \gamma/2\pi}$$

$$= \frac{(A_0/2\pi)(\gamma/2\pi)}{(f-f_0)^2 + (\gamma/2\pi)^2} - i\frac{(A_0/2\pi)(f-f_0)}{(f-f_0)^2 + (\gamma/2\pi)^2}. \tag{S3}$$

In Fig. 3d, we plot the power spectrum

$$\tilde{P}(f) = \left|\tilde{A}(f)\right|^2 = \frac{(A_0/2\pi)^2}{(f-f_0)^2 + (\gamma/2\pi)^2} \tag{S4}$$

instead of the amplitude spectrum. The full width at half maximum (FWHM) of a power spectrum peak can be deduced as

$$\tilde{P}(f_0 + f_{1/2}) = \frac{1}{2}\tilde{P}(f_0) \tag{S5}$$

$$\frac{1}{f_{1/2}^2 + (\gamma/2\pi)^2} = \frac{1}{2}\frac{1}{(\gamma/2\pi)^2} \tag{S6}$$

$$f_{\text{FWHM}} = 2f_{1/2} = \frac{\gamma}{\pi}, \tag{S7}$$

where $f_{1/2}$ is the frequency at the half maximum. Therefore, the FWHM of a power spectrum peak is given by $\gamma/\pi$ and the amplitude lifetime is $\tau = 1/\pi f_{\text{FWHM}}$. The quality factor is defined as $Q = f_0/f_{\text{FWHM}}$ and can be directly measured for each peak in Fig. 3d. The average FWHM of the phononic-cavity peaks is measured to be 4.2 GHz in the range 0.4 THz and 2.1 THz, which corresponds to the phonon amplitude lifetime of $\tau = 1/\pi(4.1 \text{ GHz}) = 78$ ps.

## 5. Reliability of fitting results

To assess the reliability of fitting results for $K_{\text{hw}}$ and $K_{\text{gh}}$ values, we repeated the fitting procedure on the time-domain data of Devices 1, 3, and 4, and the results are shown in Extended Data Fig. 6. Device 2 was excluded because it requires an additional unknown parameter, $K_{\text{ww}}$, and the exciton peak shift depends on both the hBN-WSe$_2$ and WSe$_2$-WSe$_2$ distances. We varied decay constants between devices to take into account the device-dependent defects and inhomogeneities. The best fits were obtained with similar values across the devices: $K_{\text{hw}} = 3.8 \times 10^{19}$, $4.6 \times 10^{19}$, and $3.5 \times 10^{19}$ N/m$^3$, and $K_{\text{gh}} = 7.3 \times 10^{19}$, $8.8 \times 10^{19}$, and $8.1 \times 10^{19}$ N/m$^3$ were used for Devices 1, 3, and 4, respectively. Therefore, we estimate the force constants to be $K_{\text{hw}} = (4.0 \pm 0.3) \times 10^{19}$ N/m$^3$ and $K_{\text{gh}} = (8.1 \pm 0.4) \times 10^{19}$ N/m$^3$, where the uncertainties are standard errors.

In Fig. S3, we quantify the goodness of fit for Device 4 using Pearson's linear correlation coefficient (Fig. S3a,c) and mean squared error (Fig. S3b,d) as a function of $K_{\text{hw}}$ and $K_{\text{gh}}$. The fitting results of the time-domain signals are sensitive to the value of $K_{\text{hw}}$, which determines both the reflected and transmitted signals. On the other hand, a good fit can be obtained for a wider range of $K_{\text{gh}}$ values.

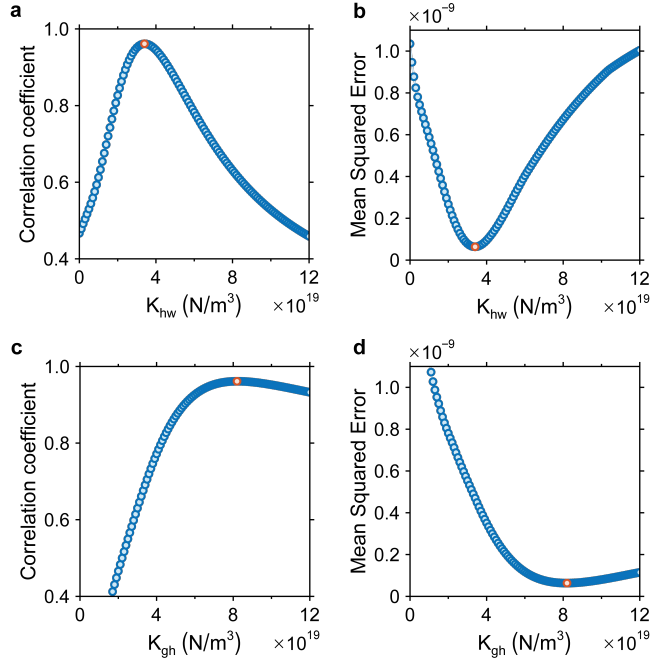

**Fig. S3 | $K_{hw}$- and $K_{gh}$-dependent fitting result of the time-domain data shown in Fig. 4d. a**, Pearson's linear correlation coefficient (**a**) and mean squared error (**b**) are plotted as a function of $K_{hw}$. The best fit is obtained with $K_{hw} = 3.5 \times 10^{19}$ N/m$^3$. The same plots (**c, d**) as a function of $K_{gh}$ shows that the best fit is obtained with $K_{gh} = 8.1 \times 10^{19}$ N/m$^3$, but a good fit can be obtained for a wider range of $K_{gh}$.

## 6. Atomic simulation of force constants

To calculate the interlayer force constants of heterointerfaces, we performed *ab initio* density-functional theory (DFT) and density-functional perturbation theory (DFPT) calculations[51]. To simulate hetero-interfaces, we constructed homo- and hetero-bilayer structures of graphene, hBN, hBN-graphene, and hBN-WSe$_2$. For graphene and hBN homo-bilayer, we adopted natural bilayer stacking configurations: AB (or Bernal) and AA' stackings for graphene and hBN bilayers, respectively. For the hBN-graphene hetero-bilayer structure we construct a 21.8°-twisted bilayer structure to mimic the lack of alignment between the layers in the experiment (Extended Data Fig. 8a). For the hBN-WSe$_2$ hetero-bilayer, we constructed a bilayer structure with a 4×4 supercell of hBN and 3×3 supercell of WSe$_2$ to account for their different lattice constants (Extended Data Fig. 8b). We relaxed the constructed bilayer structures by minimizing the DFT total energy. Subsequently, we calculated the layer-breathing mode frequencies and corresponding interlayer force constants by performing DFPT calculations.

The DFT and DFPT calculations were performed using the plane-wave Quantum ESPRESSO package[52] and using optimized norm-conserving Vanderbilt pseudopotentials[53,54]. To account for the van der Waals interactions, we adopted rVV10 nonlocal density functional[55]. Additionally, we included the calculation results using local density approximation (LDA) functional[56]. Previous studies have demonstrated that LDA functional provides more reasonable estimations of the layer-breathing-mode frequencies for multilayer graphene[57-61] and hBN[28,35,62] structures. For the semimetallic graphene and graphene-hBN calculations, we applied 0.02 Ry of Gaussian smearing for the electronic states. We ensured the convergence of the breathing mode frequency for each structure within 5 cm$^{-1}$ with respect to the *k*- point sampling and planewave energy cutoff. The DFT calculation parameters for each structure are summarized in Table S3.

In Table S4, we summarized the calculated interlayer distances, layer breathing mode frequencies, and interlayer force constants for each heterointerface. The computed force constants of graphene ($K_g$), hBN ($K_h$) and graphene-hBN ($K_{gh}$) at the DFT-LDA level are in good agreement with the measured values from this work. For hBN-WSe$_2$ interfaces, the calculated values ($K_{hw}$) are larger than the experimental values. We expect that impurities and defects buried between the hBN-WSe$_2$ layers, which were not accounted for in this study, would lower the interlayer bonding strength, thereby reducing the force constants between the two layers.

|  | $k$-point sampling | Smearing (Ry) | Plane-wave energy cutoff (Ry) |
|---|---|---|---|
| Graphene | 16 × 16 × 1 | 0.02 | 120 |
| hBN | 16 × 16 × 1 | None | 120 |
| Graphene-hBN | 4 × 4 × 1 | 0.02 | 120 |
| hBN-WSe$_2$ | 2 × 2 × 1 | None | 120 |

**Table S3 | Calculation parameters for density-functional theory and density-functional perturbation theory calculations.**

|  | DFT functional | Interlayer distance (Å) | Layer breathing mode frequency (cm$^{-1}$) | Interlayer force constant ($10^{19}$ N/m$^3$) |
|---|---|---|---|---|
| Graphene | rVV10 | 3.408 | 92.07 | 11.3 |
|  | LDA | 3.314 | 82.85 | 9.4 |
| hBN | rVV10 | 3.355 | 87.67 | 10.2 |
|  | LDA | 3.233 | 81.87 | 9.1 |
| Graphene-hBN | rVV10 | 3.429 | 91.40 | 11.1 |
|  | LDA | 3.426 | 68.83 | 6.5 |
| hBN-WSe$_2$ | rVV10 | 5.128 | 59.80 | 8.4 |
|  | LDA | 5.049 | 57.24 | 8.0 |

**Table S4 | *Ab initio* simulation result of interlayer distance, layer breathing mode frequency, and interlayer force constant.** For each structure, we performed DFT and DFPT calculations with rVV10 and LDA functional.

## References (cited in the main text)